\documentclass[twoside,11pt,letter]{article}
\usepackage[margin=1.2in]{geometry}

\usepackage{times}
\usepackage{amsmath}
\usepackage{amsfonts}
\usepackage{amssymb}
\usepackage{amsthm}
\usepackage{makeidx} 
\usepackage{graphicx}
\usepackage{mathrsfs}
\usepackage{tabularx}
\usepackage{caption}
\usepackage{bbold}
\usepackage{color}
\usepackage{fancybox}
\usepackage{verbatim}
\usepackage{hyperref}
\usepackage{tikz}
\usepackage{float}
\usepackage{url}
\usetikzlibrary{arrows,automata}
\usetikzlibrary{trees}
\usetikzlibrary{shadows}
\usepackage[framemethod=TikZ]{mdframed}
\usepackage{mdframed}
\mdfdefinestyle{theorem}{%
	linecolor=black,
	outerlinewidth=.5pt,
	roundcorner=5pt,
	innertopmargin=.5\baselineskip,
	innerbottommargin=.75\baselineskip,
	innerrightmargin=20pt,
	innerleftmargin=20pt,
	backgroundcolor=green!20,
	shadow=true,
	shadowcolor=green!20}
\mdfdefinestyle{algo}{%
	linecolor=black,
	outerlinewidth=.5pt,
	roundcorner=5pt,
	innertopmargin=.5\baselineskip,
	innerbottommargin=.75\baselineskip,
	innerrightmargin=10pt,
	innerleftmargin=10pt,
	backgroundcolor=yellow!20,
	shadow=true,
	shadowcolor=yellow!20}
\mdfdefinestyle{example}{%
	linecolor=black,
	outerlinewidth=.5pt,
	leftline = false,
	rightline = false,
	innertopmargin=.5\baselineskip,
	innerbottommargin=.75\baselineskip,
	innerrightmargin=10pt,
	innerleftmargin=10pt,
	backgroundcolor=white}

\def\bR{\mathbb{R}}

\def\b1{\mathbb{1}}

\def\eE{\mathsf{E}}

\def\eP{\mathsf{P}}

\def\tT{\mathtt{T}}

\def\fa{{\mathfrak{a}}}

\def\ff{{\mathfrak{f}}}

\def\fn{{\mathfrak{n}}}

\def\fw{{\mathfrak{w}}}
\def\fz{{\mathfrak{z}}}

\def\fI{{\mathfrak{I}}}

\def\cL{\mathcal{L}}

\def\cN{\mathcal{N}}

\def\cP{{\mathcal{P}}}
\def\cQ{{\mathcal{Q}}}

\def\cX{{\mathcal{X}}}

\def\cov{\mathsf{Cov}}

\def\var{\mathsf{Var}}

\def\e1{\mathsf{1}}

\def\of0{(0)}

\def\d{\partial}

\def\bf0{\mathbf{0}}
\def\cp1{\mathbb{CP}^1}

\theoremstyle{definition}

\theoremstyle{definition}

\begin{document}

\title{\textbf{Beyond Monte Carlo: Harnessing Diffusion Models\\ to Simulate Financial Market Dynamics}}
\author{\textbf{Andrew Lesniewski and Giulio Trigila}\\
Department of Mathematics\\
Baruch College\\
One Bernard Baruch Way\\
New York, NY 10010\\
USA}
\maketitle

\begin{abstract}
We propose a highly efficient and accurate methodology for generating synthetic financial market data using a diffusion model approach. The synthetic data produced by our methodology align closely with observed market data in several key aspects: (i) they pass the two-sample Cramer - von Mises test for portfolios of assets, and (ii) Q - Q plots demonstrate consistency across quantiles, including in the tails, between observed and generated market data. Moreover, the covariance matrices derived from a large set of synthetic market data exhibit significantly lower condition numbers compared to the estimated covariance matrices of the observed data. This property makes them suitable for use as regularized versions of the latter. For model training, we develop an efficient and fast algorithm based on numerical integration rather than Monte Carlo simulations. The methodology is tested on a large set of equity data.
\end{abstract}

\section{\label{introSec}Introduction}

In this paper, we present an efficient methodology for generating synthetic financial market data, based on the diffusion model approach. Diffusion models \cite{SWMG15}, \cite{SE19}, \cite{HJA20}, \cite{SE20}, \cite{SSKKEP21}, a class of deep generative models, are mathematical models designed to generate synthetic data by Monte Carlo simulating a reverse-time stochastic process, which is specified as an Ito stochastic differential equation (diffusion process). The diffusion model strategy to synthetic data generation and is a two stage process: encoding and decoding. This process employs the use of linear stochastic differential equations.

These models have demonstrated impressive results across various applications, including computer vision, natural language processing, time series modeling, multimodal learning, waveform signal processing, robust learning, molecular graph modeling, materials design, and inverse problem solving \cite{YZSHXZZCY23}. Despite their successes, certain aspects of diffusion models, particularly those related to the learning mechanism, require further refinement and development. Ongoing research efforts focus on addressing these performance-related challenges and enhancing the overall capabilities of diffusion modeling methodologies.

The significance of synthetic data across various domains, including addressing challenges related to data privacy, regulatory compliance, and fraud detection, is highlighted in \cite{JHCCSBMW22} and the references therein. In the context of the finance industry, the ability to generate high-quality synthetic market data plays a crucial role in applications such as portfolio allocation, portfolio and enterprise risk quantification, the design and back-testing of trading and market-making strategies, what-if analysis, and more.

In financial practice, accurately estimating various statistics of asset returns is critical. A key example is the covariance matrix, which plays a central role in quantitative metrics such as portfolio asset allocation and value at risk. These methodologies often assume, either explicitly or implicitly, that returns are drawn from a multidimensional Gaussian distribution or, at a minimum, an elliptical distribution. Various estimation techniques have been proposed, including: (i) standard sample (or maximum likelihood) estimators, (ii) shrinkage estimators \cite{LW04}, (iii) factor-based estimators \cite{FLZH20}, and others \cite{JOPSB23}. However, standard sample estimators often produce poorly conditioned covariance matrices with unstable inverses. This issue is commonly attributed to two primary factors:
\begin{itemize}
\item[(i)]{First, the dimensionality of the problem poses challenges: a covariance matrix of dimension $d$ has $d(d+1)/2$
parameters, which are estimated from a time series of $n$ observations. For typical portfolio sizes, this leads to a severely underdetermined problem. The alternative methods listed above address this issue by regularizing the sample estimator using exogenous mechanisms.}
\item[(ii)]{Second, it is often believed that the estimation process itself is somehow prone to ``error maximization'', rendering it unreliable for practical use.}
\end{itemize}

A natural approach to regularizing the estimated covariance matrix is to reliably generate a large number of samples from the unknown, non-parametric probability distribution underlying the asset returns. This would allow for the estimation of statistics, including the covariance matrix, with improved conditioning properties.

We consider a portfolio of $d$ assets $A^1,\ldots,A^d$ (e.g. equities), whose prices are observed at discrete times. By $X^1,\ldots,X^d$ we denote the returns\footnote{Defined, as usual, as $x_i=(price_{t_i} - price_{t_{i-1}})/price_{t_{i-1}.}$} on these assets. While the frequency of price observations is irrelevant to our analysis, for clarity, we assume daily observations. Notably, we impose no specific assumptions on the distribution of these returns. Denote the $d$-dimensional vectors of the asset returns at each time $t_i$ by $x_{t_i}$, $i=1,\ldots,n$, where $t_1<t_2<\ldots<t_n$. By  $\cX=\{x_i,\,i=1,\ldots,n\}$ we denote the dataset of observed values of the asset returns. The proposed method of generating synthetic data consistent with $\cX$ consists in the following steps:
\begin{itemize}
\item[1.]{We view the market data $\cX$ as a sample drawn from an unspecified probability distribution $\eP_0$. To each data point in $\cX$ random noise is added incrementally over a time interval until the data is completely deformed into white noise. This random noise is modeled as a suitably specified stochastic differential equation (SDE), chosen to smoothly transform the distribution $\eP_0$ into $d$-dimensional white noise as  $t\to1$\footnote{This choice of terminal time is conventional and does not result in any loss of generality.}. The coefficients of the SDE control how quickly the process transitions to white noise.}
\item[2.] We now run the SDE and record its terminal values at $t=1^-$. This completes the encoding part of the algorithm.
\item[3.] We train the score function of the SDE, and set up the corresponding reverse-time SDE.
\item[4.] Using the samples produced in Step 2 as terminal $t=1$ values, we generate samples from the reverse-time SDE. The $t=0$ values of these samples represent the desired synthetic market data. Importantly, there is no limit to the number of synthetic data points we can generate.  This completes the decoding part of the algorithm.
\item[5.] To assess whether the market data of size $n$ and the generated synthetic sample of size $m$ come for the same probability distribution $\eP_0$, we perform a two-sample Cramer-von Mises test.
\end{itemize}

Earlier work on this topic includes several notable publications. In \cite{KS20}, a market generator based on a restricted Boltzmann machine (RBM) was introduced. However, tests using observed market data revealed significant deviations in the tails of the RBM-generated distributions compared to the actual data. Subsequently, a generative adversarial network (GAN) methodology for generating market scenarios was developed in \cite{WKKK20}, demonstrating improved performance in the tails, aligning more closely with market data. The diffusion model generative methodology proposed in this work further enhances tail consistency, producing synthetic market scenarios that align well with observed market data.
 
From a mathematical perspective, the style of this paper is informal, avoiding formally stated theorems and proofs. Standard notations are used throughout, with the following conventions for vector operations:
\begin{itemize}
\item[1.]{We employ Dirac's notation to express various linear algebra operations, such as the inner product of two vectors $\langle x|y\rangle$, the projection operator $|x\rangle\langle x|$ on a vector, etc.}
\item[2.]{For two vectors $x,y\in\bR^d$, $x=(x^1,\ldots,x^d)$, $y=(y^1,\ldots,y^d)$ their product $xy$ is understood as the Hadamard (component-wise) product, i.e. $(xy)^i=x^i y^i$, for $i=1,\ldots,d$. Additionally, For a vector $x\in\bR^d$ and a function $f:\bR\to\bR$, $f(x)$ represents the vector whose components are the results from applying $f$ component-wise, i.e. $f(x)^i=f(x^i)$. Similarly, inequalities like $f(x)>0$ indicate that all components of $f(x)$ are positive.}
\end{itemize}

The paper is organized as follows. Section 2 reviews the methodology for model learning through score matching. For completeness, Section 3 summarizes key results from the theory of stochastic differential equations, which are applied later in the paper. Section 4 introduces the framework of denoising stochastic differential equations. Section 5 constitutes the core technical contribution, presenting an efficient algorithm for model learning via score matching. This method eliminates the need for time-consuming Monte Carlo simulations, replacing them with precise and efficient analytical approximations. In Section 6, we define the concept of synthetic market scenarios generated through diffusion modeling. Finally, Section 7 showcases a series of experiments using equity market data, validating the effectiveness of the proposed methodology.

\section{Model learning via score matching}

We begin by reviewing the score matching method and its refinement to denoising score matching. These methods are the foundations of the diffusion model approach.

\subsection{Explicit score matching}

The score matching methodology for model learning was introduced by Hyv\"arinen in \cite{H05} as a practical alternative to the maximum likelihood estimation (MLE) of the parameters of a probability distribution. The advantage of score matching over MLE is that it is directly applicable even if the probability density function (PDF) is known only up to a normalization factor. The method consists in the following.

Let $p(x)$ be the PDF of a continuous probability distribution defined for $x\in\bR^d$. Its (Stein) score $s(x)$ defined as 
\begin{equation}
s(x)=\nabla\log p(x).
\end{equation}
Notice that if $p(x)$ is defined only up to a multiplicative normalization constant (which may be hard to compute), $p(x)\sim q(x)$, its score is given by $s(x)=\nabla\log q(x)$. In particular, the score of a $d$-dimensional Gaussian distribution with mean $\mu$ and covariance $C$ is given by 
\begin{equation}\label{scoreGaussEq}
s(x)=-C^{-1}(x-\mu),
\end{equation}
an expression that does not involve the determinant of $C$. The definition of the score can easily be extended to the family of probability distribution $p(t,x)$ parameterized by time $t$.

The probability distribution $p(x)$ typically depends on a set of parameters $\theta$, i.e. $p(x)=p(x;\theta)$. The objective of model training is to learn these parameters from the available data. In the context of the Gaussian distribution, this dependence is established by specifying the functional relationship between the mean and covariance in terms of the parameters, i.e. $\mu=\mu(\theta)$, $C=C(\theta)$. 

Model learning via score matching consists in minimizing the objective function
\begin{equation}
\begin{split}
L(\theta)&=\frac12\,\eE\big(\|s(\cdot\,;\theta)-s(\cdot)\|^2\big)\\
&=\frac12\,\int_{\bR^d} \|s(x;\theta)-s(x)\|^2\, p(x)dx,
\end{split}
\end{equation}
where $p(x)$ represents here the continuum of data, and $s(x;\theta)$ is the score of the parameterized PDF\footnote{Note that the score in the sense used here is different from the Fisher score defined as $\nabla_\theta p(x;\,\theta)$.},
\begin{equation}\label{paramScEq}
s(x;\theta)=\nabla_x p(x;\theta).
\end{equation} 
Integrating by parts, one can rewrite this expression in the form \cite{H05}
\begin{equation}
L(\theta)=\int_{\bR^d}\big(\nabla^\tT s(x;\theta)+\frac12\,\|s(x;\theta)\|^2\big)\, p(x)dx+const,
\end{equation}
where $const$ is independent of $\theta$.

In practice, we use the finite sample version of this objective function. If $\cX=\{x_i,\,i=1,\ldots n\}$, $x_i\in\bR^d$, is a data sample, then the corresponding empirical objective function is given by
\begin{equation}\label{scoreObFctEq}
\begin{split}
L_\cX(\theta)&=\frac{1}{n}\,\sum_{i=1}^n\big(\nabla^\tT s(x_i;\theta)+\frac12\,\|s(x_i;\theta)\|^2\big)\\
&=\int_{\bR^d}\big(\nabla^\tT s(x;\theta)+\frac12\,\|s(x;\theta)\|^2\big)\, p_\cX(x)dx,
\end{split}
\end{equation}
where
\begin{equation}\label{empDistEq}
p_\cX(x)=\frac{1}{n}\,\sum_{i=1}^n\delta(x-x_i)
\end{equation}
is the empirical probability distribution function corresponding to the observations $\cX$. Here and in the following, $\delta(x)$ denotes the $d$-dimensional Dirac delta function. It was shown in \cite{H05} that, in case of the Gaussian distribution with $\theta$ chosen as the set of all independent parameters of $\mu$ and $C$, the minimizers of this objective function are the usual MLE values for the estimates of $\mu$ and $C$.

\subsection{\label{dsmSec}Denoising score matching}

A practically important variant of score matching, replacing the empirical distribution \eqref{empDistEq} with a smooth probability distribution, was developed by Vincent \cite{V11}. The method consists in regularizing the empirical distribution $p_\cX(x)$ by its Parzen-Rosenblatt estimate,
\begin{equation}\label{parzenEstEq}
p_{\cX,h}(x)=\frac{1}{n}\,\sum_{i=1}^n p_h(x|x_i),
\end{equation}
where $p_h(x|x_i)$ is the Gaussian kernel of bandwidth $h>0$,
\begin{equation}
p_h(x|x_i)=\frac{1}{(2\pi h^2)^{d/2}}\,e^{-\frac{1}{2h^2}\|x-x_i\|^2}.
\end{equation}
This leads to the objective function
\begin{equation}
L_{\cX,h}(\theta)=\frac12\,\int_{\bR^d} \|s(x;\,\theta)-\nabla\log p_{\cX,h}(x)\|^2\, p_{\cX,h}(x)dx,
\end{equation}
which is equivalent to \eqref{empDistEq}, as $h\to 0$. 

Interpreting $p_h(x|y)$ as a conditional probability distribution, we can express the Parzen-Rosenblatt estimate \eqref{parzenEstEq} as the margin
\begin{equation}
p_{\cX,h}(x)=\int_{\bR^d} p_h(x,y)dy
\end{equation}
of the joint distribution
\begin{equation}
p_h(x,y)=p_h(x|y)p_\cX(y).
\end{equation}
Using this, an explicit calculation \cite{V11} shows that this objective function is equivalent to the \textit{denoising score matching} (DSM) objective function given by
\begin{equation}\label{dsmEq}
L^{DSM}_{\cX,h}(\theta)=\frac12\,\int_{\bR^d} \|s(x;\,\theta)-\nabla\log p_{\cX,h}(x|y)\|^2\, p_{\cX,h}(x,y)dx\,dy.
\end{equation}

This form of the score matching principle is particularly beneficial when used with Monte Carlo simulations. We will see below that this approach is equally applicable to numerical quadrature methods. The arguments $x$ can be viewed as ``corruptions'' of the observations in the data set $\cX$ by perturbing the data by noise drawn from a suitable probability distribution. The problem of optimizing \eqref{dsmEq} can be viewed as the problem of denoising these corrupted observations.

\section{\label{dfSec}Diffusion models}

The incorporation of noise into the system will be achieved by generating sample paths of a suitable diffusion (Ito) process, which is characterized as denoising stochastic differential equations. This method facilitates the simulation of noise effects within the model.

\subsection{\label{diffProceSec}Diffusion processes}

\subsubsection{\label{diffEqs}Stochastic differential equations}

A diffusion process $X_t\in\bR^d$ over the time interval  $t\in[0,1]$ is the solution to a stochastic differential equation (SDE) \cite{O14}, \cite{IW89}
\begin{equation}\label{genSdeEq}
\begin{split}
dX_t&=\alpha(t,X_t)\,dt+\sigma(t,X_t)\,dW_t,\\
X_0&=x_0,
\end{split}
\end{equation}
where $W_t\in\bR^p$ is the standard $p$-dimensional Brownian motion. The drift vector $\alpha(t,X_t)\in\bR^d$ and diffusion matrix $\sigma(t,X_t)\in\mathrm{Mat}_{d,p}(\bR)$ are assumed to satisfy suitable technical conditions guaranteeing the existence and uniqueness of the solution. We let $p(t,x)$ denote the probability density function of $X_t$.

Note that the initial point $x_0$ is either known \textit{a priori}, or more generally, selected from an initial distribution $p_0(x)=p(0,x)$. The choice of this distribution is part of the problem specification. Within our scope, the initial distribution is assumed to be the empirical distribution \eqref{empDistEq}, $p_0(x)=p_\cX(x)$  All subsequent points $X_t$, for $t>0$, follow the distribution determined by equation \eqref{genSdeEq}.  

Let $0<u<t$, and $x,y\in\bR^d$. The transition probability $p(t,x|u, y)$ is defined as the conditional probability
\begin{equation}
p(t,x|u,y)=\eP(X_t=x|X_u=y),
\end{equation}
where $\eP$ is the probability measure associated with the process \eqref{genSdeEq}. It satisfies two fundamental partial differential equations (see the mathematics literature cited above, and also \cite{R89}, \cite{K92} for the physics perspective): 
\begin{itemize}
\item[1.]{The forward Kolmogorov equation (also known as Fokker-Planck equation) with respect to $t$ and $x$:
\begin{equation}
\begin{split}
\frac{\d}{\d t}\,p(t,x|u,y)&=-\nabla_x^\tT\big(\alpha(t,x) p(t,x|u,y)\big)+\frac12\,\nabla_x^\tT \nabla_x\big(D(t,x)p(t,x|u,y)\big),\\
p(u,x|u,y)\big)&=\delta(x-y).
\end{split}
\end{equation}}
\item[2.]{The backward Kolmogorov equation with respect $u$ and $y$:
\begin{equation}
\begin{split}
-\frac{\d}{\d u}\,p(t,x|u,y)&=\alpha(u,y)^\tT\nabla_yp(t,x|u,y)+\frac12\,D(u,y)\nabla_y^\tT\nabla_y p(t,x|u,y),\\
p(t,x|t,y)&=\delta(x-y),
\end{split}
\end{equation}
where
\begin{equation}
D(t,x)=\sigma(t,x)\sigma(t,x)^\tT
\end{equation}
is the diffusion matrix. }
\end{itemize}

Note that the probability density function $p(t,x)$ of the process $X_t$ is the solution to the forward Kolmogorov equation:
\begin{equation}\label{kolmEq}
\begin{split}
\frac{\d}{\d t}\,p(t,x)&=-\nabla_x^\tT\big(\alpha(t,x) p(t,x)\big)+\frac12\,\nabla_x^\tT \nabla_x\big(D(t,x)p(t,x)\big),\\
p(0,x)&=\delta(x-x_0).
\end{split}
\end{equation}
In the context of diffusion models, when linear SDEs are employed, the probability density function $p(t,x)$ indeed exists and is, in fact, Gaussian.

\subsubsection{Linear SDEs}

Specifically, a $d$-dimensional linear SDE driven by a $d$-dimensional Brownian motion (i.e. $p=d$ in the notation of Section \ref{genSdeEq}) is a diffusion process of the form
\begin{equation}\label{linSdeEq}
\begin{split}
dX_t&=\alpha(t, X_t)dt+\sigma(t)dW_t,\\
X_0&=x_0,
\end{split}
\end{equation}
with a linear drift
\begin{equation}
\alpha(t, X_t)=\alpha_1(t)X_t+\alpha_0(t).
\end{equation}
Here
\begin{itemize}
\item[(i)]{the coefficients $\alpha_1(t),\alpha_0(t)\in\bR^d$ are vectors of deterministic functions of time, and}
\item[(ii)]{$\sigma(t)\in\bR^d$ is a vector of positive, deterministic functions $\sigma_i(t)$.} 
\end{itemize}

The general solution to \eqref{linSdeEq} is explicitly given by
\begin{equation}\label{linEqSol}
X_t=x_0e^{\int_0^t\alpha_1(s)ds}+\int_0^t\alpha_0(v)e^{\int_v^t\alpha_1(s)ds}dv+\int_0^t\sigma(v)e^{\int_v^t\alpha_1(s)ds}\,dW_v.
\end{equation}
This is a Gaussian diffusion process whose mean is given by
\begin{equation}\label{muLinEq}
\begin{split}
\mu(t,X_0)&=\eE(X_t)\\
&=x_0e^{\int_0^t\alpha_1(s)ds}+\int_0^t\alpha_0(u)e^{\int_u^t\alpha_1(s)ds}du,
\end{split}
\end{equation}
and whose covariance matrix $C(t)$ is diagonal with
\begin{equation}\label{covLinEq}
\begin{split}
C(t)_{ij}&=\cov(X_t)_{ij}\\
&=\delta_{ij}\int_0^t\sigma_{ii}(v)^2e^{2\int_v^t\alpha_{1i}(s)ds}\,dv,
\end{split}
\end{equation}
where $\delta_{ij}$ denotes Kronecker's delta. Hence the probability law of $X_t$ is Gaussian, with density 
\begin{equation}
p(t,x)=\fn(x;\mu(t,x_0),C(t)),
\end{equation}
where $\fn(x;\mu,C)$ is the PDF of the $d$-dimensional Gaussian distribution $\cN(\mu,C)$.

More generally, it follows from \eqref{linEqSol} that, when conditioned on $X_u=y$ (where $u\leq t$), we have
\begin{equation}
X_t=y e^{\int_u^t\alpha_1(s)ds}+\int_u^t\alpha_0(v)e^{\int_v^t\alpha_1(s)ds}dv+\int_u^t\sigma(v)e^{\int_u^t\alpha_1(s)ds}\,dW_v.
\end{equation}
As a consequence, the transition probability is given by the kernel
\begin{equation}\label{transFctEq}
p(t,x|u,y)=\fn(x;\mu(t|u,y),C(t|u)),
\end{equation}
with mean
\begin{equation}
\begin{split}
\mu(t|u,y)&=\eE(X_t|X_u=y)\\
&=ye^{\int_u^t\alpha_1(s)ds}+\int_u^t\alpha_0(v)e^{\int_v^t\alpha_1(s)ds}dv,
\end{split}
\end{equation}
and covariance
\begin{equation}
\begin{split}
C(t|u)_{ij}&=\cov(X_t|X_u)_{ij}\\
&=\delta_{ij}\int_u^t\sigma_{ii}(v)^2 e^{2\int_v^t\alpha_{1i}(s)ds}\,dv.
\end{split}
\end{equation}

The multivariate SDE above is, in fact, a system of univariate SDEs. If the initial condition were drawn from a product of independent univariate distributions, studying its solution would involve solving $d$ independent one-dimensional SDEs. However, in our case, the initial condition $X_0$ is drawn from an unknown, possibly complex distribution. As a result, we consider the entire system \eqref{linEqSol} as an aggregate rather than a collection of individual univariate problems.

\subsection{\label{revTimeSec}Reverse-time diffusion process}

The key input to a diffusion model is the \textit{reverse-time diffusion process}, which corresponds to the (forward-time) diffusion process defined above. This process constitutes the decoding phase of the model, and the generated sample paths represent the synthetic scenarios. The reverse-time diffusion processes corresponding to a forward-time diffusion \eqref{genSdeEq} is an SDE which runs in reverse time and which generates the same sample paths as the original process. The general concept of reverse-time diffusion processes was introduced and studied by Anderson in \cite{A82}, who extended earlier work on reverse-time linear diffusions. 

The theory developed in \cite{A82} can be summarized as follows. Consider the diffusion process defined by \eqref{genSdeEq} with $\alpha(t,x)$ and $\sigma(t,x)$ sufficiently regular, so that the probability density function $p(t,x)$, i.e. the solution to the Kolmogorov equation \eqref{kolmEq}, exists and is unique\footnote{A general set of explicit sufficient conditions on $\alpha(t,x)$ and $\sigma(t,x)$ guaranteeing this was formulated in \cite{HP86}.}. In the case of linear diffusions, these assumptions are satisfied. We define the following $p$-dimensional process $\overline{W}_t$:
\begin{equation}
\begin{split}
d\overline{W}_t&=dW_t+\frac{1}{p(t,X_t)}\,\nabla\big(\sigma(t,X_t) p(t,X_t)\big)dt,\\
\overline{W}_0&=0.
\end{split}
\end{equation}
Furthermore, $X_t$ is independent of the increments $\overline{W}_t-\overline{W}_s$, for all $t\geq s\geq0$, and is the solution to the following SDE:
\begin{equation}\label{revTimeSdeEq}
dX_t=\overline{\alpha}(t,X_t)dt+\sigma(t,X_t)d\overline{W}_t,
\end{equation}
where
\begin{equation}\label{alphaBar}
\overline{\alpha}(t,X_t)=\alpha(t,X_t)-\frac{1}{p(t,X_t)}\,\nabla\big(D(t,X_t) p(t,X_t)\big).
\end{equation}
Equation \eqref{revTimeSdeEq} is the reverse-time SDE corresponding to \eqref{genSdeEq}. This reverse-time diffusion process mirrors the original forward-time process while running in reverse time.

It can be verified that the process $\overline{W}_t$ exhibits the natural properties of a Brownian motion running in reverse time, and is referred to as the \textit{reverse-time Brownian motion}. The presence of drift in $\overline W_t$ is, of course, related to the fact that it is associated with a different measure than the one governing $W_t$. Specifically, it is associated with the measure obtained from the original measure via Girsanov's theorem after the shift \eqref{alphaBar} of the drift term in \eqref{genSdeEq}.

In particular, if $\sigma(t,x)=\sigma(t)$ is independent of the state variable, then the formulas stated above can be expressed in terms of the score function $s(t,x)$ of the probability distribution $p(t,x)$ in a more straightforward manner. Namely,
\begin{equation}
d\overline{W}_t=dW_t+\sigma(t) s(t,x)dt,
\end{equation}
is the reverse-time Brownian motion, and
\begin{equation}
\overline{\alpha}(t,X_t)=\alpha(t,X_t)-\sigma(t)^2 s(t,x)
\end{equation}
is the reverse-time drift. The corresponding reverse-time diffusion process reads:
\begin{equation}\label{revTimeSdeEq2}
dX_t=\big(\alpha(t,X_t)-\sigma(t)^2 s(t,X_t)\big)dt+\sigma(t)d\overline{W}_t.
\end{equation}

In the case of a linear SDE of the form \eqref{linSdeEq}, the corresponding reverse-time diffusion process remains linear. This is a consequence of the explicit expression \eqref{scoreGaussEq} for the true score function, according to which $s(t, X_t)$ is linear in $X_t$. However, it should be emphasized that the the reverse-time diffusion process used for scenario generation involves the estimated score function $s(t,X_t;\theta)$, where $\theta$ denotes a collection of trainable parameters. This function is trained (e.g., using neural networks) on the data $\cX$, and it is only approximately linear. 

\section{\label{denSdeSec}Denoising SDEs and their time reversals}

A \textit{denoising SDE} (DSDE) is a linear SDE with suitable mean-reversion properties, designed to model noise within the context of a diffusion model. The following three examples are commonly used as standard realizations of DSDEs \cite{SSKKEP21}:
\begin{itemize}
\item[1.]{Variance preserving (VP) SDE.}
\item[2.]{Sub-variance preserving (sub-VP) SDE.}
\item[3.]{Variance exploding (VE) SDE.}
\end{itemize}
In the following sections, we will focus on the VP SDE. However, for completeness, we also provide brief discussions of the sub-VP and VE SDEs.

\subsection{\label{vpSec}Variance preserving SDE}

\subsubsection{Forward time VP SDE}

Let $\beta(t)\in\bR^d$ be a vector of strictly positive, deterministic functions $\beta_i(t)$, for $0<t<1$ and $i=1,\ldots, d$, such that $\int_0^t\beta_i(s)ds\to\infty$, as $t\to 1$. A convenient choice of $\beta_i(t)$ is the power function
\begin{equation}\label{betaChoiceEq}
\beta_i(t)=b_i(1-t)^{-(1+a)},
\end{equation}
where $a\geq 0$, and $b_i>0$, for $i=1,\ldots,d$, are fixed hyperparameters of the problem. These functions have the property that they grow power-like to infinity as $t\uparrow 1$.

We consider the $d$-dimensional linear SDE
\begin{equation}
dX_t=-\frac12\,\beta(t) X_t dt+\sqrt{\beta(t)}\,dW_t,
\end{equation}
According to \eqref{linEqSol}, its solution reads
\begin{equation}\label{vpSdeEq}
X_t=x_0\exp\big(-\frac12\int_0^t\beta(s)ds\big)+\int_0^t\sqrt{\beta(v)}\exp\big(-\frac12\int_v^t\beta(s)ds\big) dW_v.
\end{equation}
Thanks to the specific form of the drift and diffusion coefficients, we can express this equation more conveniently. Specifically, we introduce a new time variable:
\begin{equation}\label{tauDefEq}
\tau(t)=\int_0^t\beta(s)ds.
\end{equation}
Next, we define the time-changed process:
\begin{equation}
Z_u=X_{\tau^{-1}(u)},
\end{equation}
In other words, $Z_t$ represents the pullback of $X_t$ under the time change defined by \eqref{tauDefEq}. Leveraging the change of time formula for Brownian motion  \cite{O14},
\begin{equation}\label{timeChgEq}
\sqrt{\tau'(t)}\,dW_t=dW_{\tau(t)},
\end{equation}
and noting that $\tau(0)=0$, we can express \eqref{vpSdeEq} as follows:
\begin{equation}
Z_\tau=z_0 e^{-\tau/2}+\int_0^\tau e^{-(\tau-s)/2}\,dW_s,
\end{equation}
where $\tau=\tau(t)$ and $z_0=x_0$. 

Remarkably, the time-changed process corresponds to the multi-dimensional Ornstein-Uhlenbeck process with long-term mean zero:
\begin{equation}\label{OuSdeEq}
dZ_\tau=-\frac12\,Z_\tau d\tau+dW_\tau.
\end{equation}
The mean of $Z_\tau$ is given by
\begin{equation}\label{meanVpEq}
\eE(Z_\tau)=x_0e^{-\frac12\tau},
\end{equation}
and its variance is
\begin{equation}\label{varVpEq}
\var(Z_\tau)=1-e^{-\tau}.
\end{equation}
The PDF of the VP process is explicitly given by
\begin{equation}\label{vppEq}
p(t,x)=\fn(x;x_0 e^{-\frac12\tau(t)}, 1-e^{-\tau(t)}),
\end{equation}
which asymptotically tends to the standard normal distribution, as $t\to 1$, i.e. $X_t$ becomes white noise in this limit.

Explicitly, with the choice \eqref{betaChoiceEq}, the new time variables are given by
\begin{equation}
\tau_i(t)=
\begin{cases}
-b_i\log(1-t),&\text{ if }a=0,\\
\frac{b_i}{a}\,\big(\frac{1}{(1-t)^a}-1\big),&\text{ if }a>0.
\end{cases}
\end{equation}
The parameter $a$ determines the rate at which $\tau_i(t)$ approaches infinity as $t\uparrow 1$, with $a=0$ corresponding to the slowest rate.

\subsubsection{Reverse-time VP SDE}

According to \eqref{revTimeSdeEq2}, the reverse-time SDE for $X_t$ reads
\begin{equation}
dX_t=-\beta(t)\big(\frac12\,X_t+s(t,X_t)\big)dt+\sqrt{\beta(t)} d\overline{W}_t,
\end{equation}
where $s(t,X_t)$ is the score function of the forward-time process. As explained in Section \ref{revTimeSec}, this score function is replaced by the estimated score function. 

After time change \eqref{tauDefEq} this equation takes the form
\begin{equation}
dZ_\tau=-\big(\frac12\,Z_\tau+s(t^{-1}(\tau),Z_\tau)\big)d\tau+d\overline{W}_\tau,
\end{equation}
which is used for scenario generation.

\subsection{\label{svpSec}Sub-variance preserving SDE}

\subsubsection{Forward-time VE SDE}

The sub-variance preserving (sub-VP) SDE, introduced by \cite{SSKKEP21}, modifies the VP SDE as follows:
\begin{equation}
dX_t=-\frac12\,\beta(t)X_t dt+\sqrt{\beta(t)\big(1-e^{-2\int_0^t\beta(s)ds}\big)}\,dW_t.
\end{equation}
Rather than explicitly solving this cumbersome equation, we can directly invoke the time change method.

Using the notation introduced earlier, and applying the change of time transformation \eqref{timeChgEq}, we express this SDE in the form:
\begin{equation}
dZ_\tau=-\frac12\,Z_\tau d\tau+\sqrt{1-e^{-2\tau}}\,dW_\tau.
\end{equation}
The solution for $Z_\tau$ is given by:
\begin{equation}
Z_\tau=z_0 e^{-\tau/2}+\int_0^{\tau}\sqrt{1-e^{-2s}}\,e^{-(\tau-s)/2}\,dW_s.
\end{equation}
While the mean of $Z_\tau$ remains given by \eqref{meanVpEq}, its variance becomes:
\begin{equation}
\var(Z_\tau)=(1-e^{-\tau})^2.
\end{equation}
This expression is dominated by \eqref{varVpEq}. 

The PDF of the sub-VP process is explicitly given by
\begin{equation}
p(t,x)=\fn(x;x_0 e^{-\tau(t)}, (1-e^{-\tau(t)})^2),
\end{equation}
Consequently, $X_t$ tends to pure white noise, as $t\to 1$, albeit with a lower variance compared to that of the VE SDE.

\subsubsection{Reverse-time sub-VP SDE}

According to \eqref{revTimeSdeEq}, the reverse-time SDE for $X_t$ reads
\begin{equation}
dX_t=-\beta(t)\big(\frac12\,X_t+\big(1-e^{-2\int_0^t\beta(s)ds}\big)s(t,X_t)\big)dt+\sqrt{\beta(t)\big(1-e^{-2\int_0^t\beta(s)ds}\big)}\, d\overline{W}_t,
\end{equation}
where $s(t,X_t)$ is the (estimated) score function. After time change \eqref{tauDefEq} this equation takes the form:
\begin{equation}
dZ_\tau=-\big(\frac12\,Z_\tau+\big(1-e^{-2\tau}\big)s(t^{-1}(\tau),Z_\tau)\big)d\tau+d\overline{W}_\tau.
\end{equation}

\subsection{\label{veSec}Variance exploding SDE}

\subsubsection{Forward-time VE SDE}

Let $v(t)\in\bR^d$ be a vector of positive, strictly increasing differentiable functions. A convenient choice is
\begin{equation}
v_i(t)=b_i a^t,
\end{equation}
with constant hyperparameters $a>1$ and $b_i>0$, for $i=1,\ldots,d$.

We consider the following linear, driftless SDE:
\begin{equation}\label{varExpEq}
dX_t=\sqrt{\frac{d}{dt}\,v(t)}\,dW_t
\end{equation}
It's solution is given by
\begin{equation}
X_t=X_0+\int_0^t \sqrt{\frac{d}{ds}\,v(s)}\,dW_s,
\end{equation}
and so
\begin{equation}
\eE(X_t)=x_0,
\end{equation}
and
\begin{equation}
\cov(X_t)=v(t).
\end{equation}

This solution can also be understood in terms of change of time formula \eqref{timeChgEq} with
\begin{equation}
\tau(t)=v(t).
\end{equation}
Explicitly,
\begin{equation}
dZ_\tau=dW_\tau,
\end{equation}
with $Z_0=z_0(=x_0)$, and 
\begin{equation}
\tau(t)=b a^t.
\end{equation}

Clearly, the probability density function of \eqref{varExpEq} is explicitly given by
\begin{equation}
p(t,x)=\fn(x;x_0,\tau(t)),
\end{equation}
which explodes to infinity, as $t\to\infty$.

\subsubsection{Reverse-time VE SDE}

According to \eqref{revTimeSdeEq}, the reverse-time SDE for $X_t$ reads
\begin{equation}
dX_t=-s(t,X_t)\,\frac{d}{dt}\,v(t)\,dt+\sqrt{\frac{d}{dt}\,v(t)}\,d\overline{W}_t,
\end{equation}
After change of time this equation takes the form
\begin{equation}
dZ_\tau=- s(t^{-1}(\tau),Z_\tau)d\tau+d\overline{W}_\tau.
\end{equation}

\section{\label{trainSec}Diffusion model training via score matching}

The parameters $\theta$ are estimated from the objective function \cite{SSKKEP21}
\begin{equation}
\cL^{SM}(\theta)=\frac12\,\int_0^1\int_{\bR^d}\lambda(t)\|s(t,x;\theta)-\log p(t,x)\|^2 p(t,x)\,dx\,dt,
\end{equation}
where $\lambda(t)>0$ is a suitably chosen weight function. Its purpose is to regularize the behavior of $\|s(t,x;\theta)-\log p(t,x)\|^2$ at small values of $t$. As explained in Section \ref{dsmSec}, this objective function is equivalent to the DSM objective function \cite{V11}:
\begin{equation}
\cL^{DSM}(\theta)=\frac12\,\int_0^1\int_{\bR^d\times\bR^d}\lambda(t)\|s(t,x;\theta)-\log p(t,x|0,x_0)\|^2 p(t,x|0,x_0)p_0(x_0)\,dx_0\,dx\,dt,
\end{equation}
where $p(t,x|0,x_0)=p(t,x)$ is the transition probability density \eqref{vppEq}, and $p_0(x_0)$ is the (empirical) distribution of the data set $\cX$ \eqref{empDistEq}. 

\subsection{Parameterization of the DSM objective function}

According to \eqref{scoreGaussEq}, the model score function $s(t,x)$ is given by
\begin{equation}\label{scoreExplEq}
s(t,x)=-\frac{1}{C(t)}\,\big(x-\mu(t, x_0)\big),
\end{equation}
where we have taken advantage of the fact that our covariance matrix is diagonal. We parameterize the fitted score $s(t,x;\theta)$ as a fully connected neural net with a single hidden layer, namely
\begin{equation}\label{fitScoreFct}
s(t,x;\theta)=\frac{1}{C(t)}\,K(x;\theta).
\end{equation}
Here $K(x;\theta)$ is explicitly given by
\begin{equation}
K^i(x;\theta)=\sum_{j=1}^h c^{ij}\fa\big(\sum_{k=1}^d w_{jk} x^k+b_j\big)+d^i,
\end{equation}
where $i=1,\ldots, d$, and where $\fa(u)$ is a smooth, positive activation function, such as the softplus function, 
\begin{equation}
\fa(u)=\log(1+e^u).
\end{equation}
The neuronal weights and biases $\theta=(w, b, c, d)$ have the following properties: 
\begin{itemize}
\item[(i)]{$w\in\mathrm{Mat}_{h,d}(\bR)$, where $h$ is the number of neurons in the hidden layer,}
\item[(ii)]{$b\in\bR^h$,}
\item[(iii)]{$c\in\mathrm{Mat}_{d,h}(\bR)$, and}
\item[(iv)]{$d\in\bR^d$.}
\end{itemize} 
Notice that we use upper and lower indices in accordance with our convention to label the components of a single $x$ with upper indices, while examples of $x$ are labeled with lower indices. We refer to this network as the noise conditional score network (NCSN).

Using \eqref{transFctEq} and \eqref{scoreExplEq} to compute the transition probability, we can rewrite this expression in the form
\begin{equation}
\begin{split}
\cL^{DSM}(\theta)&=\frac12\,\int_0^1\int_{\bR^d\times\bR^d}\lambda(t)\big\|s(t,x;\theta)+\frac{x-\mu(t,x_0)}{C(t)}\big\|^2\, p(t,x|0,x_0)p_0(x_0)\,dx_0\,dx\,dt\\
&=\frac{1}{2n}\sum_{i=1}^n\int_0^1\int_{\bR^d}\lambda(t)\big\|\frac{K(x;\theta)+x_i-\mu(t,x_i)}{C(t)}\big\|^2\,\ff^i_t(x)dx\,dt,
\end{split}
\end{equation}
where we have introduced the notation
\begin{equation}\label{ffDef}
\ff^i_t(x)=\fn(x;\mu(t,x_i),C(t)).
\end{equation}

We will now choose $\lambda(t)$ to have the form:
\begin{equation}\label{lambdaEq}
\lambda(t)=\lambda_0(t)C(t)^2,
\end{equation}
where $\lambda_0(t)$ is continuous for all $t\geq 0$, and so
\begin{equation}
\cL^{DSM}(\theta)=\frac{1}{2n}\sum_{i=1}^n\int_0^1\int_{\bR^d}\lambda_0(t)\|K(x;\theta)+x_i-\mu(t,x_i)\|^2\,\ff^i_t(x)dx\,dt.
\end{equation}

Explicitly, in case of the VP and sub-VP denoising diffusions, this expression reads explicitly
\begin{equation}\label{vpDsmEq}
\cL^{DSM}(\theta)=\frac{1}{2n}\sum_{i=1}^n\int_0^1\int_{\bR^d}\lambda_0(t)\|K(x;\theta)+x_i-x_ie^{-\tau(t)}\|^2\, \ff^i_t(x)dx\,dt,
\end{equation}
with the appropriate covariances defined in Sections \ref{vpSec} and \ref{svpSec}, respectively. Analogous expression for the VE denosing SDE reads
\begin{equation}\label{veDsmEq}
\cL^{DSM}(\theta)=\frac{1}{2n}\sum_{i=1}^n\int_0^1\int_{\bR^d}\lambda_0(t)\|K(x;\theta)+x_i\|^2\, \ff_t^i(x)dx\,dt,
\end{equation}
with the covariance defined in Section \ref{veSec}.

\subsection{\label{integSec}Evaluation of the DSM objective function}

The high-dimensional integrals in \eqref{vpDsmEq} and \eqref{veDsmEq} necessitate an efficient numerical evaluation method. Integration over $t$ is carried out accurately using Simpson's rule. In fact, a small number of subdivisions of the integral over $[0,1]$ (such as 10 or fewer) is typically sufficient to achieve an accurate result. Integration over $x$ can, in principle, be carried out via Monte Carlo simulations. This method requires very significant computing resources even if the dimensionality $d$ of the problem is moderate, and is impractical for large values of $d$ (such as $d>100$). However, owing to the specific form of the objective function, the $d$-dimensional integration can be reduced to one- and two-dimensional Gaussian integrals which can be efficiently computed using Gauss - Hermite quadrature.

We start by expressing $\cL^{DSM}(\theta)$ as a sum:
\begin{equation*}
\begin{split}
\cL^{DSM}(\theta)=&\frac{1}{2n}\sum_{i=1}^n\int_0^1\lambda_0(t)\int_{\bR^d}\|K(x;\theta)+x_i-\mu(t,x_i)\|^2\,\ff^i_t(x)dx\,dt\\
=&\frac{1}{2n}\,\sum_{i=1}^n\sum_{k=1}^d \int_0^1 \lambda_0(t)(x^k_i-\mu(t,x_i)^k+d^k)^2 dt\\
&+\frac{1}{n}\,\sum_{i=1}^n\sum_{j=1}^h\sum_{k=1}^d c^{jk}\int_0^1\lambda_0(t)(x_i^k-\mu(t,x_i)^k+d^k)\int_{\bR^d}\fa(\langle w_j|x\rangle +b_j)\,\ff^i_t(x)dx\,dt\\
&+\frac{1}{2n}\,\sum_{i=1}^n\sum_{j,k=1}^h c^{ij} c^{ik}\int_0^1\lambda_0(t)\int_{\bR^d}\fa(\langle w_j|x\rangle+b_j)\fa(\langle w_k|x\rangle+b_k)\,\ff^i_t(x)dx\,dt,
\end{split}
\end{equation*}
where $\ff^i_t(x)$ is given by \eqref{ffDef}, and where $w_j,w_k\in\bR^d$ denote the $j$-th and $k$-th rows of the matrix $w$, respectively. Notice that, with probability one, $w_j$ and $w_k$, are linearly independent for $j\neq k$. In order to organize the computation, we rewrite this expression as follows:
\begin{equation}\label{Ldsm_expl}
\begin{split}
\cL^{DSM}(\theta)=&\frac{1}{2n}\,\sum_{i=1}^n\sum_{k=1}^d \int_0^1 \lambda_0(t)(x^k_i-\mu(t,x_i)^k+d^k)^2 dt\\
&+\frac{1}{n}\,\sum_{i=1}^n\sum_{j=1}^h\sum_{k=1}^d c^{jk}\int_0^1\lambda_0(t)(x_i^k-\mu(t,x_i)^k+d^k)\fI_1^i(w_j, b_j, t)\,dt\\
&+\frac{1}{2n}\,\sum_{i=1}^n\sum_{j,k=1}^h c^{ij} c^{ik}\int_0^1\lambda_0(t)\fI^i_2(w_j, w_k,b_j, b_k,t)\,dt,
\end{split}
\end{equation}
where
\begin{equation}
\fI_1^i(w,b,t)=\int_{\bR^d}\fa\big(\langle w|x\rangle+b\big)\ff_t^i(x)dx,
\end{equation}
and
\begin{equation}
\fI^i_2(w_1, w_2,b_1, b_2,t)=\int_{\bR^d}\fa\big(\langle w_1|x\rangle+b_1\big)\fa\big(\langle w_2|x\rangle+b_2\big)\ff_t^i(x) dx,
\end{equation}
for linearly independent $w_1$ and $w_2$. Changing variables we express $\fI_1$ and $\fI_2$ as integrals with respect to the standard $d$-dimensional Gaussian measure: 
\begin{equation}
\fI_1^i(w,b,t)=\frac{1}{(2\pi)^{d/2}}\,\int_{\bR^d}\fa\big(\langle w|\sqrt{C(t)}\,x+\mu(t,x_i)\rangle+b\big)e^{-\frac12\langle x|x\rangle}\, dx,
\end{equation}
and
\begin{equation}
\begin{split}
\fI^i_2(w_1, w_2,b_1, b_2,t)=\frac{1}{(2\pi)^{d/2}}\,&\int_{\bR^d}\fa\big(\langle w_1|\sqrt{C(t)}\,x+\mu(t,x_i)\rangle+b_1\big)\\
&\times\fa\big(\langle w_2|\sqrt{C(t)}\,x+\mu(t,x_i)\rangle+b_2\big)e^{-\frac12\langle x|x\rangle}\, dx,
\end{split}
\end{equation}
respectively.

The $d$-dimensional integrals $\fI_1$ are $\fI_2$ are actually highly numerically tractable and can be reduced to one- and two-dimensional Gaussian integrals, respectively. In order to see it, we set
\begin{equation}
E=\frac{1}{\|w\|}\,w,
\end{equation}
and rewrite $\fI_1^i(w,b,t)$ as
\begin{equation}
\fI_1^i(w,b,t)=\frac{1}{(2\pi)^{d/2}}\,\int_{\bR^d}\fa\big(\|w\|\sqrt{C(t)}\langle E|x\rangle+\langle w|\mu(t,x_i)\rangle+b\big) e^{-\frac12\langle x|x\rangle}\,dx.
\end{equation}
Let $y=\langle E|x\rangle$ and $\Gamma=I-|E\rangle\langle E|$. Then
\begin{equation}
\begin{split}
\fI_1^i(w,b,t)&=\int_{\bR^d}\fa\big(\|w\|\sqrt{C(t)}\,y+\langle w|\mu(t,x_i)\rangle+b\big)
e^{-\frac12\,y^2-\frac12\langle x|\Gamma x\rangle}dy\, d^{d-1}x\\
&=\frac{1}{(2\pi)^{1/2}}\int_{\bR}\fa\big(\|w\|\sqrt{C(t)}\,y+\langle w|\mu(t,x_i)\rangle+b\big)e^{-\frac12\,y^2}dy,
\end{split}
\end{equation}
as $\det(\Gamma)=1$. This integral can be efficiently evaluated using the Gauss-Hermite quadrature algorithm,
\begin{equation}
\fI_1^i(w,b,t)\approx\sum_{p=1}^D\fw_p\fa\big(\|w\|\sqrt{2C(t)}\,\fz_p+\langle w|\mu(t,x_i)\rangle+b\big).
\end{equation}
Here $\fz_p$, $p=1,\ldots,D$, are the zeros of the Hermite polynomial of degree $D$, and $\fw_p$ are the appropriate weights. Even low values of $D$, such as $D=4$, are sufficient to calculate the integral accurately. Note the presence of $\sqrt{2}$ above, arising from the weight function in the integral being $e^{-\frac12\, y^2}$ rather than $e^{-y^2}$.

Similarly, let us define the normalized vectors
\begin{equation}
E_i=\frac{1}{\|w_i\|}\,w_i,
\end{equation}
for $i=1,2$. We use the Gram-Schmidt algorithm to orthonormalize these vectors, which yields
\begin{equation}
\begin{split}
E_1'&=E_1,\\
E_2'&=\frac{1}{\sqrt{1-\langle E_2|E_1\rangle^2}}\,(E_2-\langle E_2|E_1\rangle E_1).
\end{split}
\end{equation}
Let $y_i=\langle E_i'|x\rangle$, for $i=1,2$, and $\Gamma=I-|E_1'\rangle\langle E_1'|-|E_2'\rangle\langle E_2'|$. We also let 
\begin{equation}
\mathrm{c}(w_1,w_2)=\frac{\langle w_1|w_2\rangle}{\|w_1\|\|w_2\|}\,,
\end{equation}
and
\begin{equation}
\mathrm{s}(w_1,w_2)=\sqrt{1-\frac{\langle w_1|w_2\rangle^2}{\|w_1\|^2\|w_2\|^2}}
\end{equation}
denote the cosine and (the absolute value of the) sine of the angle between $w_1$ and $w_2$, respectively. Then
\begin{equation*}
\begin{split}
\fI_2^i&(w_1,w_2)\\
&=\frac{1}{(2\pi)^{d/2}}\int_{\bR^d}\fa\big(\|w_1\|\sqrt{C(t)}\,y_1+\langle w_1|\mu(t,x_i)\rangle+b_1\big)\\
&\quad\times\fa\big(\|w_2\|\sqrt{C(t)}\,(\mathrm{s}(w_1,w_2)y_2+\mathrm{c}(w_1,w_2)y_1)+\langle w_2|\mu(t,x_i)\rangle+b_2\big) e^{-\frac12\,(y_1^2+y_2^2)-\frac12\langle x|\Gamma x\rangle}d^2y\, d^{d-2}x\\
&=\frac{1}{2\pi}\int_{\bR^2}\fa\big(\|w_1\|\sqrt{C(t)}\,y_1+\langle w_1|\mu(t,x_i)\rangle+b_1\big)\\
&\qquad\qquad\times\fa\big(\|w_2\|\sqrt{C(t)}\,(\mathrm{s}(w_1,w_2)y_2+\mathrm{c}(w_1,w_2)y_1)+\langle w_2|\mu(t,x_i)\rangle+b_2\big) e^{-\frac12\,(y_1^2+y_2^2)}d^2y,
\end{split}
\end{equation*}
as $\det(\Gamma)=1$. This two-dimensional Gaussian integral can be efficiently evaluated using the Gauss-Hermite quadrature algorithm,
\begin{equation*}
\begin{split}
\fI_2^i(w_1,w_2)&\approx\sum_{1\leq p,q\leq D}\fw_p\fw_q\fa\big(\|w_1\|\sqrt{2C(t)}\,\fz_p+\langle w_1|\mu(t,x_i)\rangle+b_1\big)\\
&\quad\times\fa\big(\|w_2\|\sqrt{2C(t)}\,(\mathrm{s}(w_1,w_2)\fz_q+\mathrm{c}(w_1,w_2)\fz_p)+\langle w_2|\mu(t,x_i)\rangle+b_2\big),
\end{split}
\end{equation*}
with a low value of $D$.

Finally, as explained at the beginning of this section, we perform time integration using Simpson's method. As a result of these calculations, we obtain an expression for \eqref{Ldsm_expl} that is suitable for implementation in TensorFlow, enabling fast and accurate computations.

\section{\label{synthMktSec}Synthetic markets}

The process of generating synthetic markets involves several tasks:
\begin{itemize}
\item[(i)]{Training of the fitted score function, as discussed in Section \ref{trainSec}.}
\item[(ii)]{Encoding and decoding of the training (historically observed) data.}
\item[(iii)]{Testing whether the generate data follow the same probability distribution as the training data. This step is crucial, as financial data are often difficult to visualize.}
\end{itemize}
In this section we focus on tasks (ii) and (iii).

\subsection{\label{eulerSec}Euler-Maruyama scheme for the forward and reverse-time DSDE}

The process of generating a synthetic market scenario involves two runs of the denoising SDE: a forward run and a reverse-time run. Both runs effectively draw a sample from the probability distribution associated with a DSDE. This is achieved using the Euler-Maruyama numerical scheme \cite{K92}, which simulates approximate sample paths of the SDE.

\noindent
\textit{Forward path generation.} To generate a forward path for the linear SDE \eqref{linSdeEq} starting at  $X_{init}$, we divide the interval $[0, 1]$ into $K+1$ equal subintervals of length $\delta=1/K$ using the points $t_j=j/K$, where $j=0,\ldots, K$, and denote by $\hat X_{t_j}$ the (approximate) value of $X_t$ at $t=t_j$.
\begin{itemize}
\item[1.]{We start at the initial time $t=t_0=0$, and set
\begin{equation}
\hat X_{t_0}=X_{init}.
\end{equation}}
\item[2.]{Next, for $j=1,2,\ldots,K$, simulate increments for a $d$-dimensional Brownian motion $d\hat{W}_j\sim\cN(0,\delta I_d)$, and set
\begin{equation}
\hat X_{t_{j+1}}=\hat X_{t_j}+\alpha(t_j,\hat X_{t_j})\delta+\sigma(t_j)d\hat{W}_j.
\end{equation}}
\end{itemize}

\noindent
\textit{Reverse-time path generation.} To simulate sample paths of the reverse-time SDE \eqref{revTimeSdeEq2} starting at $X_{term}$, an appropriate score function $s(t, x)$ is required. In practice, we utilize the fitted score function \eqref{fitScoreFct}, defining $\hat s(t,x)=s(t,x,\theta)$.

As with forward path generation, we divide the interval $[0, 1]$ into $K+1$ equal subintervals of length $\delta=1/K$, denoting the division points as $t_j=j/K$, where $j=0,\ldots, K$. Let $\hat X_{t_j}$ represent the (approximate) value of $X_t$ at $t=t_j$. The procedure is as follows:
\begin{itemize}
\item[1.]{We start at the terminal time $t=t_K=1$, and set
\begin{equation}
\hat X_{t_K}=X_{term}.
\end{equation}}
\item[2.]{Next, for $j=K,K-1,\ldots,1$, simulate increments for a $d$-dimensional Brownian motion $d\hat{W}_j\sim\cN(0,\delta I_d)$. Then, update using the following equations:
\begin{equation}
\begin{split}
d\hat{\overline W}_j&=d\hat W_j+\sigma(t_j) s(t_j,\hat X_{t_j})\delta,\\
\hat X_{t_{j-1}}&=\hat X_{t_j}+\big(\alpha(t_j,\hat X_{t_j})-\sigma(t_j)^2 s(t_j,\hat X_{t_j})\big)\delta+\sigma(t_j)d\hat{\overline W}_j.
\end{split}
\end{equation}}
\end{itemize}
Note: The Brownian motion used for reverse-time path generation should be freshly simulated and must not reuse the Brownian motion from forward path generation. Reusing it would merely recover the initial point, without the advantage of introducing noise.

\subsection{\label{genMktSec}Generating synthetic market scenarios}
Given a training set of market data $\cX$ representing $n$ observations of returns on $d$ assets, we aim at generating $m$ synthetic returns on these assets, where the choice of $m$ is unrelated to $n$. We proceed as follows:
\begin{itemize}
\item[A.]{Train the fitted score function \eqref{fitScoreFct} on the dataset $\cX$, as explained in Section \ref{trainSec}.}
\item[B.]{Generate synthetic samples. For $k=1,\ldots,m$ perform the following steps:
\begin{itemize}
\item[(i)]{Randomly select an index $i_k\in\{1,\ldots,n\}$.}
\item[(ii)]{Set $X^k_{init}=x_{i_k}$, and generate a forward path starting at $X^k_{init}$. Let $\hat X^k_1$ denote the terminal value of this path.}
\item[(iii)]{Set $X^k_{term}=\hat X^k_1$, and generate a reverse-time path starting at $X^k_{term}$. Let $\hat x_k$ denote the initial ($t=0$) value of this path.}
\end{itemize}}
\item[C.]{The set $\hat\cX=\{\hat x_1,\ldots,\hat x_m\}$ is the desired set of $m$ synthetic market scenarios.}
\end{itemize}

It remains to verify that $\hat\cX$ is drawn from the same distribution as the original market data $\cX$. Multivariate tests for assessing the goodness of fit between two empirical distributions - see, for example, \cite{KBW20} - are notoriously challenging and difficult to implement. In a forthcoming paper \cite{LT24}, we propose a practical methodology that we believe is well-suited for situations like the one addressed in this study.

In the meantime, we adapt the univariate two-sample Cramer-von Mises (CvM) criterion \cite{A62} to our context. To this end, we consider a specified portfolio of assets $A^1,\ldots,A^d$ by choosing their weights $g=(g_1,\ldots,g_d)$, where $g_i\geq 0$, for all $i$, and $\sum_{i=1}^d g_i=1$. Let
\begin{equation}
\begin{split}
p_i&=\langle g|x_i\rangle,\quad\text{ for } i=1,\dots,n,\\
\hat q_k&=\langle g|\hat x_k\rangle,\quad\text{ for } k=1,\dots,m,\\
\end{split}
\end{equation}
denote the returns on this portfolio under the historical and synthetic scenarios, respectively. Let $\cP=\{p_1,\ldots,p_n\}$ and $\hat{\cQ}=\{\hat q_1,\ldots,\hat q_m\}$ denote the two datasets of portfolio returns.

The two-sample CvM criterion tests the hypothesis that the two samples $\cP$ and $\hat{\cQ}$ are drawn from the same distribution. This is done using the test statistics $T_{CvM}$ \cite{A62}, whose value depends of course on the choice of weights $g$. The null hypothesis, which posits that the two samples come from the same distribution (i.e. $T_{CvM}=0$), can be rejected at a desired level of confidence.

While no choice of the portfolio weights $g$ is \textit{a priori} preferred, numerical simulations \cite{LT24} suggest that an equally weighted portfolio $g=(1/d,\ldots,1/d)$ often lies near the maximizer (i.e. the worst case scenario) of $T_{CvM}$. We use the corresponding $p$-value of the CvM test, denoted in the following by $p_{CvM}$, as a provisional metric for the significance level of the multivariate test assessing whether the datasets $\cX$ and $\hat{\cX}$ are drawn from the same distribution.
 
\section{\label{expSec}Experiments}

For our experiments, we downloaded daily closing prices from Bloomberg\footnote{The Bloomberg closing prices are adjusted for stock splits but not for dividends. We believe this does not impact our simulation methodology} for 33 large U.S. equities over the five-year period from November 13, 2019, to November 13, 2024.\footnote{AAPL, AXP, BA, CAT, COST,	CSCO, CVX, DD, 	DIS, GE, GS, HD, IBM, INTC,	JNJ, JPM, KO, LLY, MCD, MMM, MRK, MSFT, NIKE, NVDA, PFE, PG, TRV, TSLA, UNH, V, VZ, WMT, XOM.}. This period notably encompasses the economic distress caused by the COVID-19 pandemic, which introduced significant volatility to equity markets. Testing the model’s ability to accurately simulate under these non-stationary conditions serves as a critical assessment of its validity.

We selected various data windows, each $n=256$ days in length - approximately equivalent to the number of trading days in a year. For each window, we trained the diffusion model on the corresponding data and generated $m=1,024$ synthetic market scenarios. 

The model was configured as follows: the VP DSDE was selected with instantaneous volatility functions defined by \eqref{betaChoiceEq}, using parameters $a=0$ and $b_i=0.1$ for all $i$. Gauss-Hermite integration was performed with an order of $D=4$, and Simpson integration was performed using 8 subintervals. The function $\lambda_0(t)$ in \eqref{lambdaEq} was set to $\lambda_0(t)=1$, and the number of hidden neurons was configured as $h=16$. Training was conducted using the Adam optimizer.

To evaluate the model’s performance for each run, we recorded the following metrics:
\begin{itemize}
\item[(i)] The p-value $p_{CvM}$ of the CvM statistic of the equally weighted portfolio of all selected assets.
\item[(ii)] The condition number $\kappa_{hist}$ of the covariance matrix of the synthetic market data compared to that of the historical data, $\kappa_{synth}$.
\item[(iii)] The Q-Q plot comparing the synthetic data to the historical market data for visual assessment.
\end{itemize}

\noindent
\textit{Experiment 1}. We use days 1 through 256 of the data set. Historically, this period covered the Covid pandemic, highly elevated market volatility, and aggressive rate cuts by the Federal Reserve.
\begin{equation}
\begin{split}
p_{CvM}&=1.00,\\
\kappa_{hist}&=109.64,\\
\kappa_{synth}&=53.46.
\end{split}
\end{equation}

\begin{figure}[H]
\centering
\begin{minipage}{0.45\textwidth}
\centering
\includegraphics[width=\textwidth]{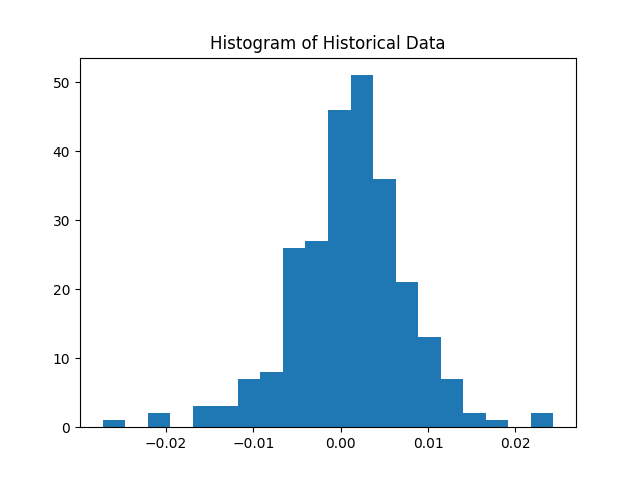}
\end{minipage}
\hfill
\begin{minipage}{0.45\textwidth}
\centering
\includegraphics[width=\textwidth]{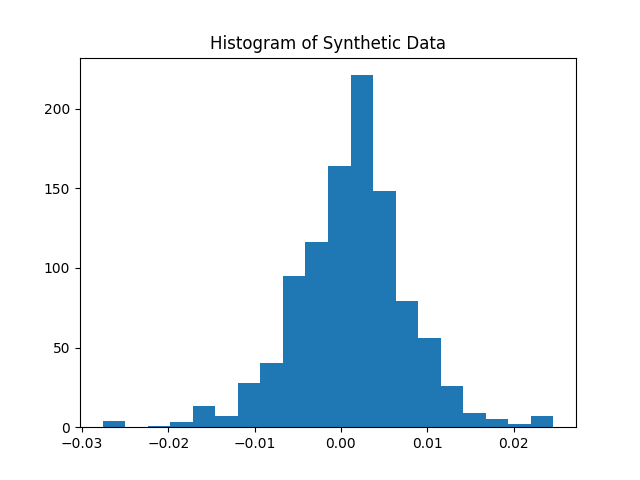}
\end{minipage}
\caption{\small Histograms of historical (left) and synthetic (right) returns in Experiment 1.}
\label{fig:hist_1}
\end{figure}

\begin{figure}[H]
\centering
\scalebox{1.0}[1.0]{\includegraphics[height=6.6cm]{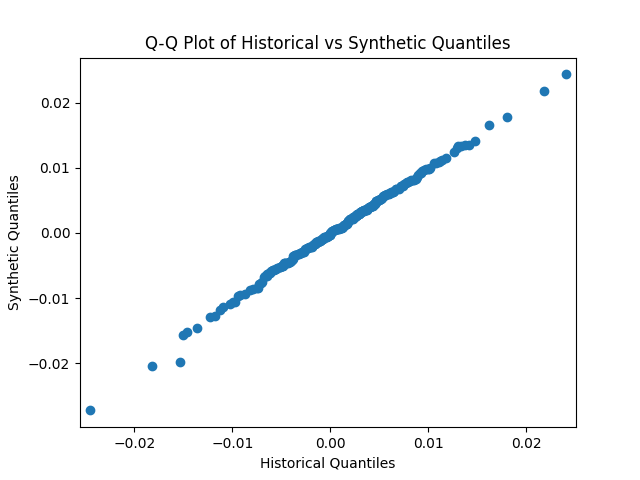}}
\caption{\small Q-Q plot of the historical versus synthetic returns in Experiment 1.}
\label{fig:qq_1}
\end{figure}

\noindent
\textit{Experiment 2}. We use days 250 through 506 of the data set. This period was marked by historically low interest rates and low volatility of equities.
\begin{equation}
\begin{split}
p_{CvM}&=0.64,\\
\kappa_{hist}&=130.37,\\
\kappa_{synth}&=70.63.
\end{split}
\end{equation}

\begin{figure}[H]
\centering
\begin{minipage}{0.45\textwidth}
\centering
\includegraphics[width=\textwidth]{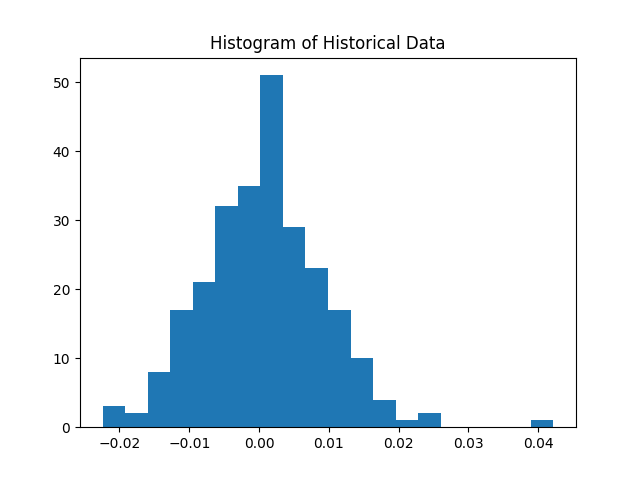}
\end{minipage}
\hfill
\begin{minipage}{0.45\textwidth}
\centering
\includegraphics[width=\textwidth]{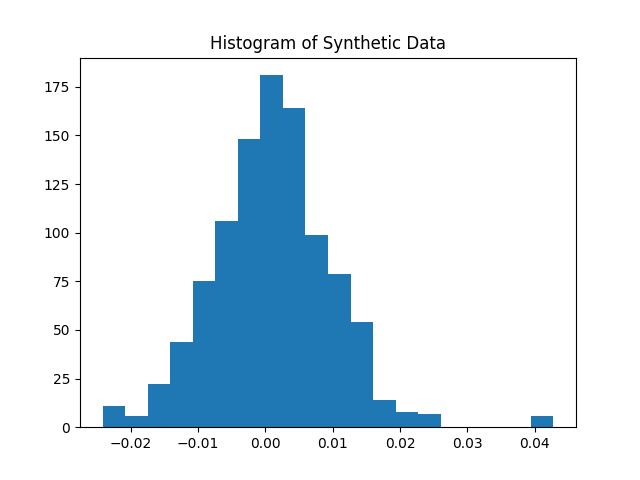}
\end{minipage}
\caption{\small Histograms of historical (left) and synthetic (right) returns in Experiment 2.}
\label{fig:hist_2}
\end{figure}

\begin{figure}[H]
\centering
\scalebox{1.0}[1.0]{\includegraphics[height=6.6cm]{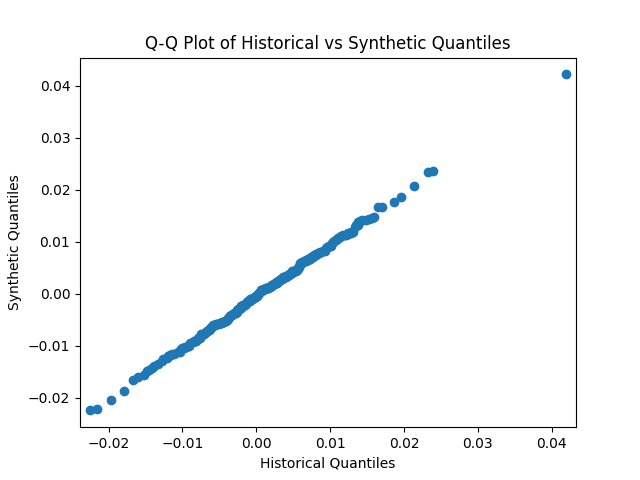}}
\caption{\small Q-Q plot of the historical versus synthetic returns in Experiment 2.}
\label{fig:qq_2}
\end{figure}

\noindent
\textit{Experiment 3}. We use days 500 through 756 of the data set. This period was marked by rising inflation, elevated market volatility, and interest rate hikes implemented by the Federal Reserve.
\begin{equation}
\begin{split}
p_{CvM}&=1.00,\\
\kappa_{hist}&=208.17,\\
\kappa_{synth}&=124.40.
\end{split}
\end{equation}

\begin{figure}[H]
\centering
\begin{minipage}{0.45\textwidth}
\centering
\includegraphics[width=\textwidth]{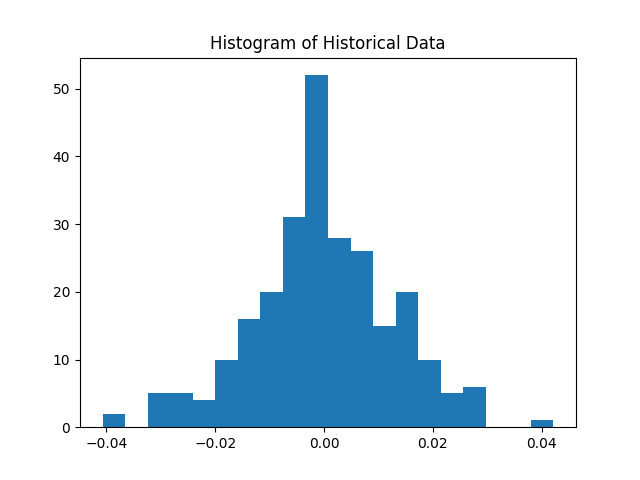}
\end{minipage}
\hfill
\begin{minipage}{0.45\textwidth}
\centering
\includegraphics[width=\textwidth]{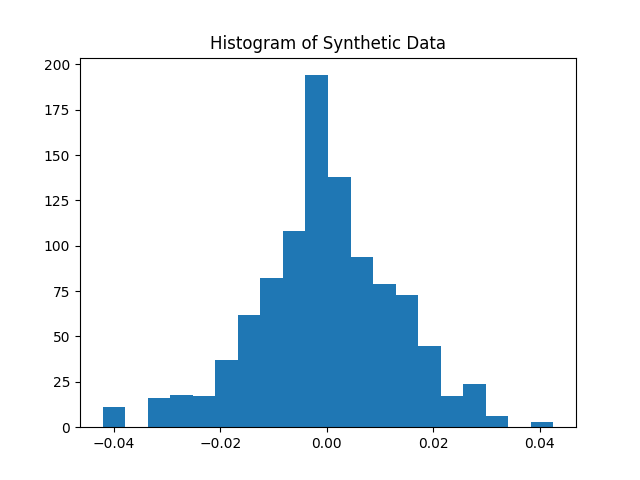}
\end{minipage}
\caption{\small Histograms of historical (left) and synthetic (right) returns in Experiment 3.}
\label{fig:hist_3}
\end{figure}

\begin{figure}[H]
\centering
\scalebox{1.0}[1.0]{\includegraphics[height=6.6cm]{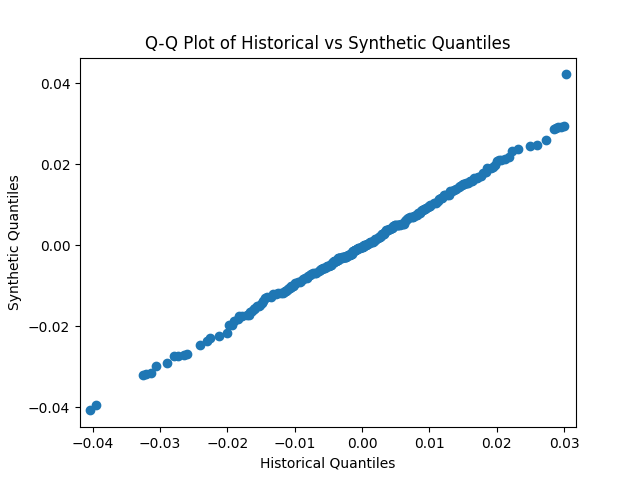}}
\caption{\small Q-Q plot of the historical versus synthetic returns in Experiment 3.}
\label{fig:qq_3}
\end{figure}

\noindent
\textit{Experiment 4}. We use days 750 through 1006 of the data set. During this period, the Federal Reserve concluded its rate-hiking program.
\begin{equation}
\begin{split}
p_{CvM}&=0.83,\\
\kappa_{hist}&=95.82,\\
\kappa_{synth}&=51.05.
\end{split}
\end{equation}

\begin{figure}[H]
\centering
\begin{minipage}{0.45\textwidth}
\centering
\includegraphics[width=\textwidth]{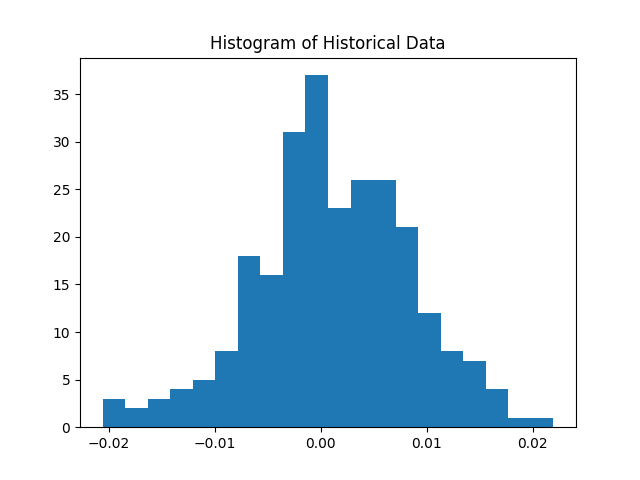}
\end{minipage}
\hfill
\begin{minipage}{0.45\textwidth}
\centering
\includegraphics[width=\textwidth]{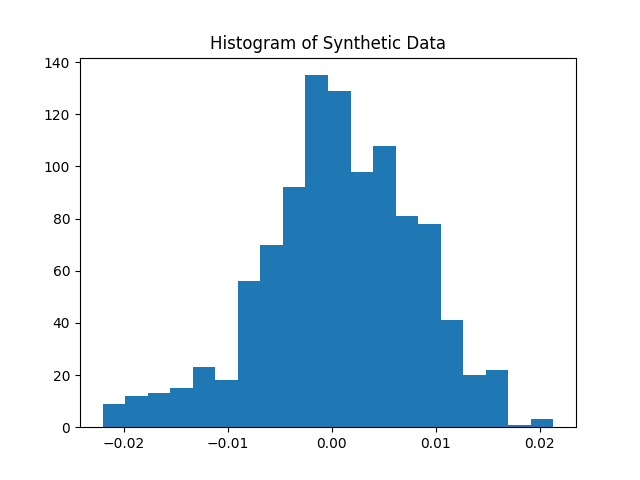}
\end{minipage}
\caption{\small Histograms of historical (left) and synthetic (right) returns in Experiment 4.}
\label{fig:hist_4}
\end{figure}

\begin{figure}[H]
\centering
\scalebox{1.0}[1.0]{\includegraphics[height=6.6cm]{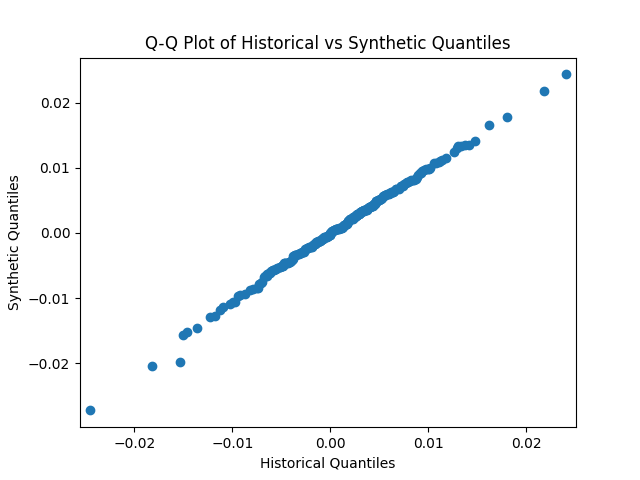}}
\caption{\small Q-Q plot of the historical versus synthetic returns in Experiment 4.}
\label{fig:qq_4}
\end{figure}

\noindent
\textit{Experiment 5}. We use days 1000 through 1256 of the data set. This period was characterized by occasional spikes in market volatility driven by the presidential election.
\begin{equation}
\begin{split}
p_{CvM}&=0.98,\\
\kappa_{hist}&=438.51,\\
\kappa_{synth}&=282.80.
\end{split}
\end{equation}

\begin{figure}[H]
\centering
\begin{minipage}{0.45\textwidth}
\centering
\includegraphics[width=\textwidth]{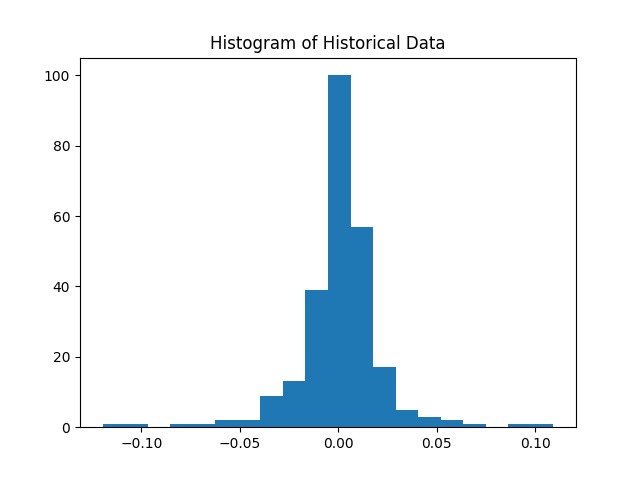}
\end{minipage}
\hfill
\begin{minipage}{0.45\textwidth}
\centering
\includegraphics[width=\textwidth]{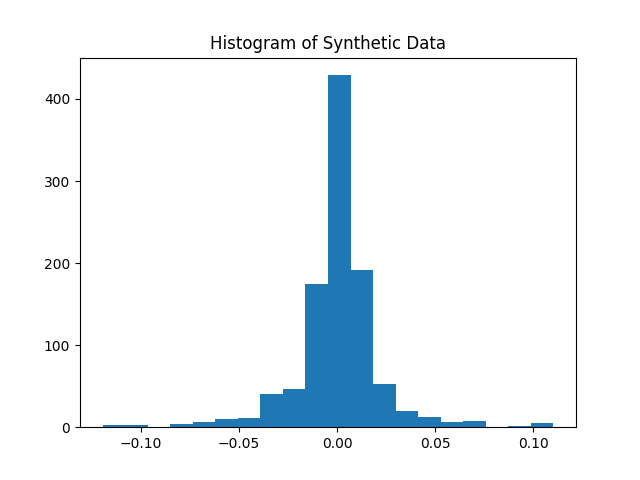}
\end{minipage}
\caption{\small Histograms of historical (left) and synthetic (right) returns in Experiment 5.}
\label{fig:hist_5}
\end{figure}

\begin{figure}[H]
\centering
\scalebox{1.0}[1.0]{\includegraphics[height=6.6cm]{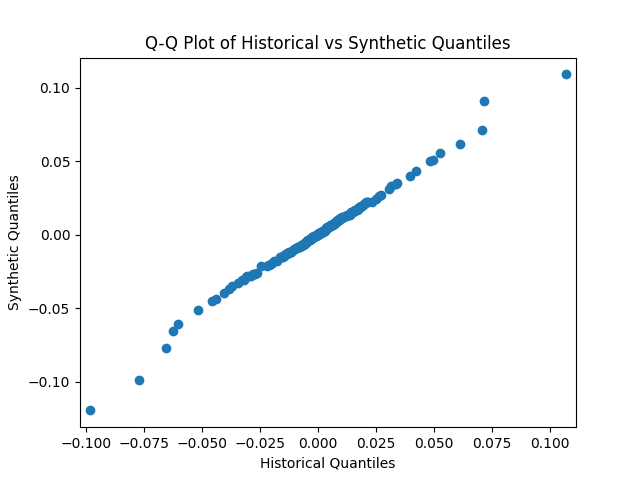}}
\caption{\small Q-Q plot of the historical versus synthetic returns in Experiment 5.}
\label{fig:qq_5}
\end{figure}

\noindent
\textit{Experiment 6}. In our final experiment we use days 1 through 1248 of the data set, which covers the entire five year period\footnote{We use 1248 (rather than 1258) for convenience, as 1248 is divisible by 32, our batch size parameter.}. In order to accommodate a larger data set, for this experiment we increased the number of hidden neurons to $h=32$.
\begin{equation}
\begin{split}
p_{CvM}&=0.99,\\
\kappa_{hist}&=130.56,\\
\kappa_{synth}&=85.03.
\end{split}
\end{equation}

\begin{figure}[H]
\centering
\begin{minipage}{0.45\textwidth}
\centering
\includegraphics[width=\textwidth]{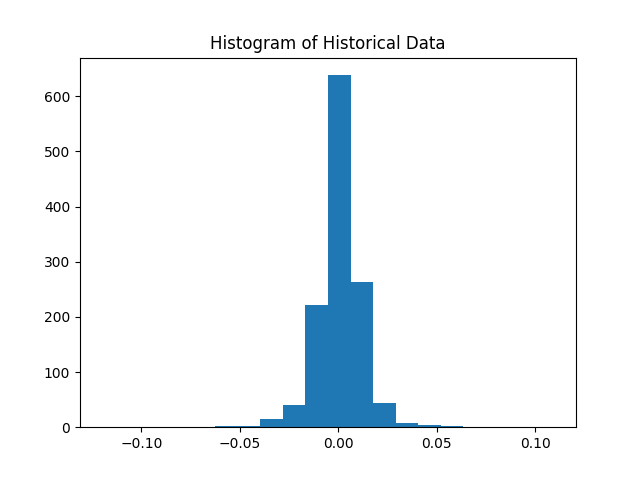}
\end{minipage}
\hfill
\begin{minipage}{0.45\textwidth}
\centering
\includegraphics[width=\textwidth]{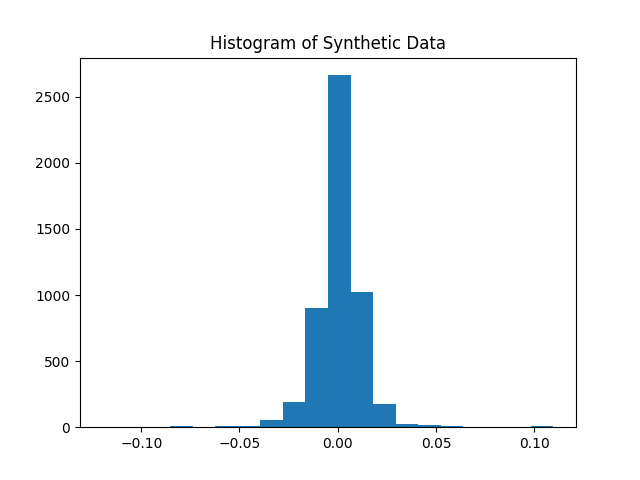}
\end{minipage}
\caption{\small Histograms of historical (left) and synthetic (right) returns in Experiment 6.}
\label{fig:hist_6}
\end{figure}

\begin{figure}[H]
\centering
\scalebox{1.0}[1.0]{\includegraphics[height=6.6cm]{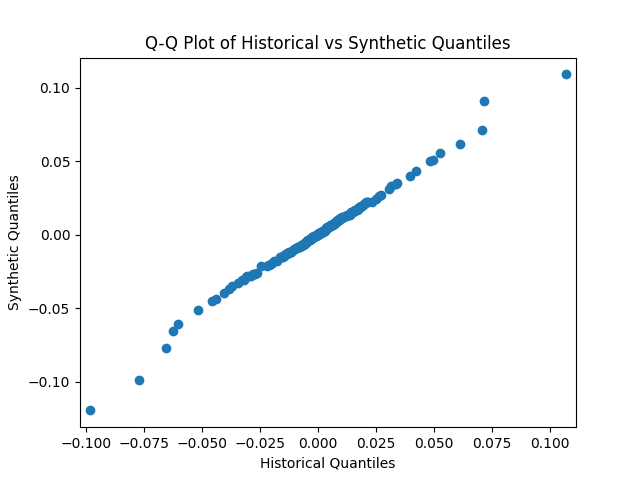}}
\caption{\small Q-Q plot of the historical versus synthetic returns in Experiment 6.}
\label{fig:qq_6}
\end{figure}

It is worth noting that the period used for our experiments was characterized by rapidly shifting economic regimes, including market volatility from the pandemic, falling interest rates, subsequent inflation, rising rates, and political uncertainty surrounding the presidential election. Despite these challenges, the model consistently demonstrated strong performance across the different time windows.

The model can generate an arbitrary number of synthetic data points without compromising their tail behavior or the CvM $p$-value. Moreover, the condition number of the estimated covariance matrix decreases as the sample size grows. For instance, in Experiment 1, where 16,384 synthetic scenarios were generated, the covariance matrix condition number was $\kappa_{synth}=47.33$, compared to $\kappa_{synth}=53.46$ for 1,024 synthetic scenarios.

\section{Conclusion and future work}

In this work, we introduced a novel method for estimating the score function within diffusion models to generate synthetic financial market scenarios. By leveraging Gauss-Hermite quadrature instead of conventional Monte Carlo estimation, our approach effectively captures the distribution of asset returns, particularly its tails, which are often characterized by sparse data points. This capability makes the method especially well-suited for applications in financial modeling. A detailed study of its application to portfolio optimization will be presented in a separate paper.

This study serves as a proof of concept for the proposed algorithm. While we employed a two-layer neural network for parameterization, this choice represents just one of many possible configurations. Future research will focus on expanding this work along two primary directions.

The first direction involves utilizing neural networks with more complex architectures, introducing additional layers of parameters into the expression for $\cL^{DSM}(\theta)$, as described at the beginning of Section \ref{integSec}. Depending on the specific application, a few additional layers may suffice to more accurately capture higher-order moments beyond the covariance structure analyzed in this study.

The second approach investigates enriching the parameterization of the score function through a “gradient flow” methodology, as outlined in Section 1.1 of \cite{TT13} in the context of Normalizing Flows. In cases where a two-layer neural network is insufficient to parameterize the score function mapping the empirical measure $\rho_{0}(x) = p(t=0, x)$ to a standard normal distribution, the method described in this work can be applied iteratively. In this framework, the push forward measure $\rho_{1}(x) = p(t=1, x)$ from the previous step becomes the starting measure $\rho_{1} = p_{1}(t=0, x)$ for the subsequent step. This iterative process generates a sequence of measures, $\rho_{0} \to \rho_{1} \to \dots \to \mathcal{N}(0, I)$, that converges to the standard normal distribution. This approach approximates the overall score function required to map the initial measure $\rho_{0}$ to $\cN(0, I)$ as a composition of mappings, each derived from a score function parameterized with only two layers.

\end{document}